\documentclass[11pt]{article}
\usepackage{cospar}
\usepackage{url}
\usepackage[dvips]{graphicx}
\usepackage[english]{babel}

\title{GRB~980425, SN1998BW AND THE EMBH MODEL}

\author{R. Ruffini\address{ICRA - International Centre for Relativistic Astrophysics and Dipartimento di Fisica, Universit\`a di Roma ``La Sapienza'', Piazzale Aldo Moro 5, I-00185 Roma, Italy}, M.G. Bernardini$^{1}$, C.L. Bianco$^{1}$, P. Chardonnet\address{Universit\'e de Savoie, LAPTH - LAPP, BP 110, F-74941 Annecy-le-Vieux Cedex, France}, F. Fraschetti\address{Universit\`a di Trento, Via Sommarive 14, I-38050 Povo (Trento), Italy.}, S.-S. Xue$^{1}$}

\begin{document}

\maketitle

\begin{abstract}
The EMBH model, previously developed using GRB~991216 as a prototype, is here applied to GRB~980425. We fit the luminosity observed in the 40--700 keV, 2--26 keV and 2--10 keV bands by the BeppoSAX satellite. In addition we present a novel scenario in which the supernova SN1998bw is the outcome of an ``induced gravitational collapse'' triggered by GRB~980425, in agreement with the GRB-Supernova Time Sequence (GSTS) paradigm (Ruffini et al. 2001c). A further outcome of this astrophysically exceptional sequence of events is the formation of a young neutron star generated by the SN1998bw event. A coordinated observational activity is recommended to further enlighten the underlying scenario of this most unique astrophysical system.
\end{abstract}

\section*{INTRODUCTION}

The aim of this talk is to present the application of the EMBH theory, previously successfully applied to GRB~991216 used as a prototype, to the case of GRB~980425 (Pian et al., 2000) and SN1998bw (Galama et al., 1998). This is a particularly important test for the validity of the EMBH theory over a range of energies of 6 orders of magnitude: as we will see, both sources appear to be spherically symmetric and the respective total energies are $E_{tot}\simeq 5\times 10^{53}\, {\rm ergs}$ and $E_{tot}\simeq 10^{48}\, {\rm ergs}$. We recall that the EMBH theory (see Ruffini et al., 2003b) depends only on three parameters, the energy of the dyadosphere $E_{dya}$, the $B$ parameter and the factor ${\cal R}$ describing the interstellar medium (ISM) porosity. The theory, therefore, explains all the observed features of the bolometric intensity variations of the afterglow as well as the spectral properties of the source and, in the specific case of GRB~980425, it also allows to clarify the general astrophysical scenario in which the GRB actually occurs. In this system, in fact, we propose that GRB~980425 has been the trigger of a phenomenon of ``induced gravitational collapse'' (Ruffini et al., 2001c) originating the supernova explosion and we also witness the birth of a young neutron star out of the supernova event. This extraordinary coincidence of these three astrophysical events represents an unprecedented scenario of fundamental importance in the field of relativistic astrophysics. Using the EMBH theory we shall explore: {\bf a)} the process of black hole formation in the event GRB~980425 (Pian et al., 2000), {\bf b)} the concept of ``induced gravitational collapse'', introduced in the GRB-Supernova Time Sequence (GSTS) paradigm (Ruffini et al., 2001c), and its link to a very special supernova type in SN1998bw (Galama et al., 1998), and finally {\bf c)} the observation for the first time of the cooling of a hot newly formed neutron star.

The observational situation of this system is quite complex. In addition to the source GRB~980425 and the supernova SN1998bw, two X-ray sources have been found by BeppoSAX in the error box for the location of GRB~980425: a source {\em S1} and a source {\em S2} (Pian et al., 2000). Since the nature of the two sources S1 and S2 was not clear, a variety of slopes in the decaying part of the afterglow have been proposed (see Fig.~1). Kulkarni et al. (1998) have proposed to explain both the supernova SN1998bw and the GRB~980425 observations by a new class of GRBs, distinctly different from the cosmological ones, both originated by a single unusual supernova event. Similarly, Iwamoto et al. (1998) have tried to explain both the supernova event and the GRB with a new kind of supernova with an extremely large explosion energy, larger than $10^{52}\, {\rm ergs}$, which they identify with the ``hypernovae'' predicted by Paczy\'{n}ski (1998). In this approach a totally novel concept is introduced: the supernova itself is assumed to originate in the process of gravitational collapse to a black hole of a massive progenitor star ($\sim 40M_\odot$) with a particularly large angular momentum and strong magnetic field. A large rotational energy of the black hole extracted with a strong magnetic field is called in, by these authors, to explain the successful explosion of this ``hypernova'' leading both to the GRB and the supernova.

\begin{figure}
\centering
\includegraphics[width=\hsize]{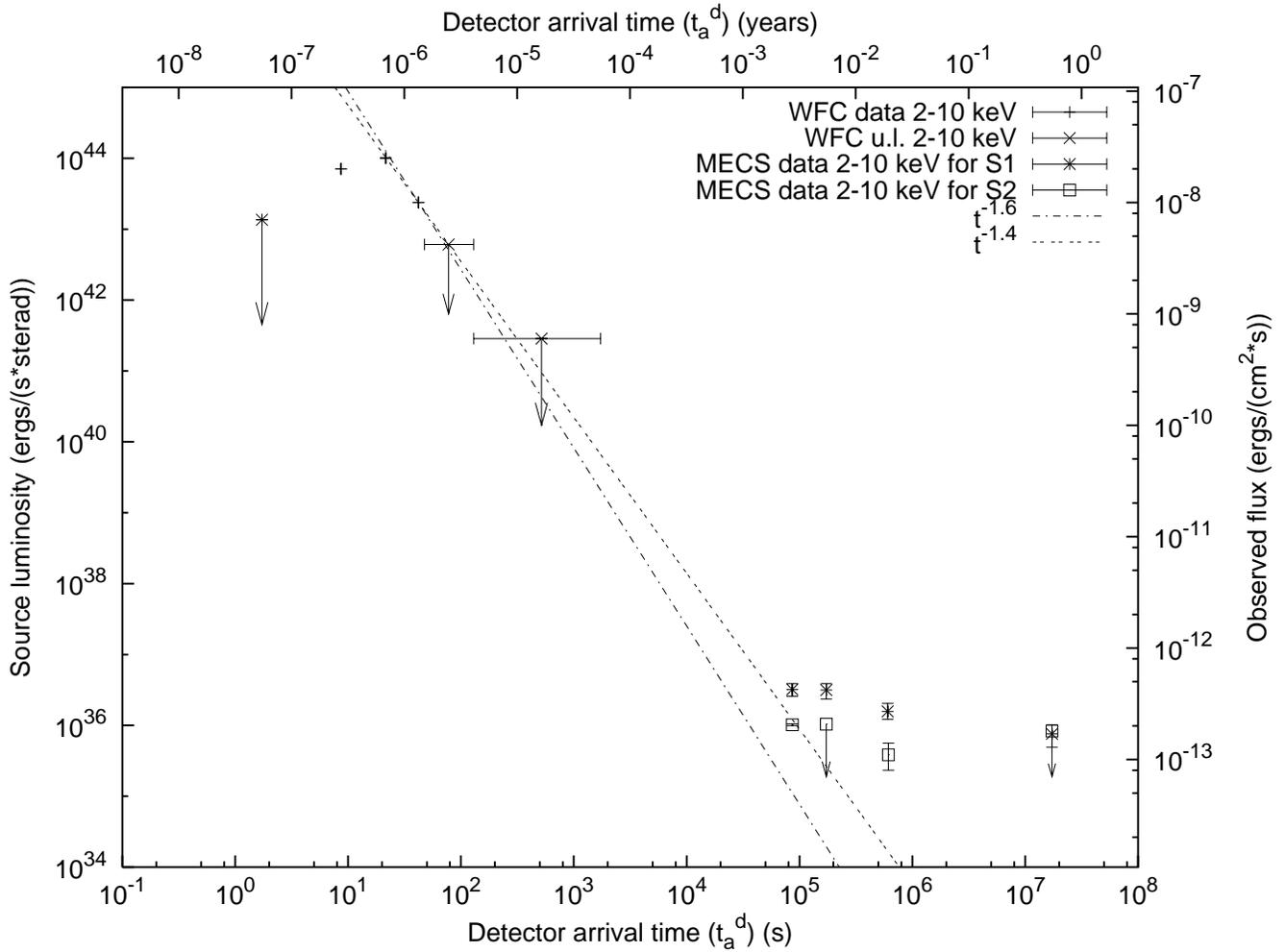}
\caption{The BeppoSAX MECS light curves in the 2-10 keV band of the X-ray sources S1 and S2 from Pian et al. (2000) detected in the GRB~980425 fields are shown. The WFC early measurement and $3\sigma$ upper limits in the same band are also shown. The dotted lines represent the power laws of indexes $\simeq 1.4$ and $\simeq 1.6$ connecting the last WFC measurement and the first and last of the first 5 MECS points of S2 which are in this picture replaced by the first point that has been obtained by integrating and averaging the flux in shorter time interval, as in Pian et al. (2000). In the vertical axis is reported (left) the luminosity of the source and (right) the observed flux as a function of the arrival time at the detector in seconds (bottom) and in years (top).}
\end{figure}

Our approach is drastically different. We first interpret the GRB~980425 within the EMBH theory. This allows the computation of the luminosity, spectra, Lorentz gamma factors, and more generally all the dynamical aspects of the source. Having characterized the features of GRB~980425, we can gradually approach the remaining part of the scenario, disentangling the GRB observations from the supernova ones and from the sources S1 and S2. This leads to a natural time sequence of events and to their autonomous astrophysical characterization.

\section*{THE ENERGETICS, DYNAMICAL PARAMETERS AND SPACE-TIME PARAMETRIZATION OF GRB~980425}

Our approach has focused on identifying the energy extraction process from the black hole (Christodoulou \& Ruffini, 1971) as the basic energy source for the GRB phenomenon. The distinguishing feature is a theoretically predicted source energetics all the way up to $1.8\times 10^{54}\left(M_{BH}/M_{\odot}\right) {\rm ergs}$ for $3.2 M_{\odot} \le M_{BH} \le 7.2 \times 10^6 M_{\odot}$ (Damour \& Ruffini, 1975). In particular, the formation of a ``dyadosphere'', during the gravitational collapse leading to a black hole endowed with electromagnetic structure (EMBH) has been indicated as the initial boundary conditions of the GRB process (Ruffini, 1998; Preparata et al., 1998). Our model has been referred as ``the EMBH model'', although the EMBH physics only determines the initial boundary conditions of the GRB process by specifying the physical parameters and spatial extension of the neutral electron positron plasma originating the phenomenon created in the dyadosphere. The creation of this plasma is due to the vacuum polarization process occurring in a supercritical field by the Heisenberg-Euler-Schwinger process (see Heisenberg \& Euler, 1935; Schwinger, 1951; Damour \& Ruffini, 1975; Preparata et al., 1998).

Traditionally, following the observations of the {\em Vela} (Strong, 1975) and {\em CGRO}\footnote{see http://cossc.gsfc.nasa.gov/batse/} satellites, GRBs have been characterized by few parameters such as the fluence, the characteristic duration ($T_{90}$ or $T_{50}$) and the global time averaged spectral distribution (Band et al., 1993). With the observations of {\em BeppoSAX}\footnote{see http://www.asdc.asi.it/bepposax/} and the discovery of the afterglow, and the consequent optical identification, the distance of the GRB source has been determined and consequently the total energetics of the source has been added as a crucial parameter.

The observed energetics of GRBs, computed for spherically symmetric explosions, do coincide with the ones theoretically predicted in Damour \& Ruffini (1975). This has been the major reason which has motivated us to reconsider and develop in full details the EMBH model. For simplicity, we have considered the vacuum polarization process occurring in an already formed Reissner-Nordstr\"om black hole (Ruffini, 1998; Preparata et al., 1998), whose dyadosphere has an energy $E_{dya}$. It is clear, however, that this is only an approximation to the real dynamical description of the process of gravitational collapse to an EMBH. In order to prepare the background for attacking this extremely complex dynamical process, we have clarified some basic theoretical issues, necessary to implement the description of the fully dynamical process of gravitational collapse to an EMBH (see Ruffini \& Vitagliano, 2002,2003; Cherubini et al., 2002).

We have then given the constitutive equations for the five eras in the EMBH model (see for details Ruffini et al., 2003a and references therein). {\em The Era I}: the $e^+e^-$ pairs plasma, initially at $\gamma=1$, self propels itself away from the dyadosphere as a sharp pulse (the PEM pulse), reaching Lorentz gamma factor of the order of 100 (Ruffini et al., 1999). {\em The Era II}: the PEM pulse, still optically thick, engulfs the remnant left over in the process of gravitational collapse of the progenitor star with a drastic reduction of the gamma factor; the mass $M_B$ of this engulfed baryonic material is expressed by the dimensionless parameter $B=M_Bc^2/E_{dya}$ (Ruffini et al., 2000). {\em The Era III}: the newly formed pair-electromagnetic-baryonic (PEMB) pulse, composed of $e^+e^-$ pair and of the electrons and baryons of the engulfed material, self-propels itself outward reaching in some sources Lorentz gamma factors of $10^3$--$10^4$; this era stops when the transparency condition is reached and the emission of the proper-GRB (P-GRB) occurs (Bianco et al., 2001). {\em The Era IV}: the resulting accelerated baryonic matter (ABM) pulse, ballistically expanding after the transparency condition has been reached, collides at ultrarelativistic velocities with the baryons and electrons of the interstellar matter (ISM) which is assumed to have an average constant number density, giving origin to the afterglow. {\em The Era V}: this era represents the transition from the ultrarelativistic regime to the relativistic and then to the non relativistic ones (see Ruffini et al., 2003a,2003b and references therein).

The EMBH model differs in many respects from the models in the current literature. The major difference consists in the following points:\\
{\bf a)} The appropriate theoretical description of all the above mentioned five eras is implemented, as well as the evaluation of the process of vacuum polarization originating the dyadosphere. The description of the inner engine originating the GRBs has never been addressed in the necessary details in the literature;\\
{\bf b)} The dynamical equations as well as the description of the phenomenon in the laboratory time and the time sequence carried by light signals recorded at the detector have been explicitly integrated (see e.g. Ruffini et al., 2003a, 2003b). In doing so we have also corrected a basic conceptual inadequacy, common to all the current works on GRBs, which led to an improper spacetime parametrization of the GRB phenomenon, preempting all these works from their predictive power: the relation between the photon arrival time at the detector and their emission time in the laboratory frame, expressed in our approach by an integral of a function of the Lorentz gamma factor extended over all the past source worldlines, has been in the current literature expressed as a function of an instantaneous value of the Lorentz gamma factor. These two approaches are conceptually very different and lead to significant qualitative differences (Ruffini et al., 2003a, 2003b and references therein).\\
{\bf c)} The treatment of the afterglow has been widely considered in the literature by the so-called ``fireball model'' (see e.g. M\'esz\'aros \& Rees, 1992, 1993; Rees \& M\'esz\'aros, 1994; Piran, 1999 and references therein). However, also in the description of the afterglow, there are major differences between the works in the literature and our approach (Ruffini et al., 2003b).

The equations of motion in our model depend only on two free parameters: the total energy $E_{tot}$, which coincides with the dyadosphere energy $E_{dya}$, and the amount $M_B$ of baryonic matter left over from the gravitational collapse of the progenitor star, which is determined by the dimensionless parameter $B=M_Bc^2/E_{dya}$. The best fit of GRB~980425 is reproduced in Tab.~1. It correspond to $E_{dya}=1.1\times 10^{48}\, {\rm ergs}$, $B=7\times 10^{-3}$ and the ISM average density is found to be $\left<n_{ism}\right>=0.02 {\rm particle}/{\rm cm}^3$. The plasma temperature and the total number of pairs in the dyadosphere are respectively $T=1.028\, {\rm MeV}$ and $N_{e^\pm}=5.3274\times10^{53}$.

\begin{table}[t]
\centering
\caption{Gamma factor for selected events and their space-time coordinates: $r$ is the radial coordinate in the laboratory frame, $t$ is the laboratory time and $t_{a}^{d}$ is the photon arrival time at the detector. The points $1,2,3,4$ mark the beginning of first four ``eras'' (see Ruffini et al., 2003a,2003b) until the transparency point (point $4$). The points $A,B,C$ correspond to three events respectively before, during and after the peak of the E-APE. The points $D,E,F,G,H,I,J$ are taken in the decaying part of the afterglow. The last column shows how the apparent motion in the radial coordinate, evaluated in the arrival time at the detector, leads to an enormous apparent ``superluminal'' behaviour. This illustrates the impossibility of using such a classical estimate in regimes of high gamma Lorentz factor.}
\begin{tabular}{l|l|l|l|l|l}
\multicolumn{6}{c}{ }\\
 Point & $r(cm)$ & $t(s)$ & $t_{a}^{d}(s)$ & $\gamma$ & ``Superluminal'' $v\equiv\frac{r}{t_{a}^{d}}$\\
\hline
$1$ & $8.79\times 10^{6}$ & $2.60\times10^{-5}$ & $2.03\times10^{-5}$ & $1.027$ & $14c$\\
$2$ & $2.89\times 10^{8}$ & $9.56\times 10^{-3}$ & $2.13\times10^{-4}$ & $28.423$ & $47c$\\
$3$ & $3.45\times 10^{8}$ & $1.14\times 10^{-2}$ & $2.19\times10^{-4}$ & $8.647$ & $53c$\\
$4$ & $4.31\times 10^{11}$ & $14.4$ & $7.65\times10^{-4}$ & $138.863$ & $1.9\times10^{4}c$\\
\hline
$A$ & $4.01\times 10^{15}$ & $1.34\times 10^{5}$ & $3.65$ & $129.819$ & $3.7\times10^{4}c$\\
$B$ & $6.00\times 10^{15}$ & $2.00\times 10^{5}$ & $5.87$ & $115.309$ & $3.4\times10^{4}c$\\
$C$ & $6.50\times 10^{15}$ & $2.17\times 10^{5}$ & $6.98$ & $84.704$ & $3.1\times10^{4}c$\\
\hline
$D$ & $1.02\times 10^{16}$ & $3.40\times 10^{5}$ & $19.2$ & $53.982$ & $1.8\times10^{4}c$\\
$E$ & $1.07\times 10^{16}$ & $3.57\times 10^{5}$ & $98.5$ & $6.510$ & $3.6\times10^{3}c$\\
$F$ & $1.12\times 10^{16}$ & $3.74\times 10^{5}$ & $3.00\times10^{2}$ & $6.469$ & $1.2\times10^{3}c$\\
$G$ & $1.17\times 10^{16}$ & $3.90\times 10^{5}$ & $5.00\times10^{2}$ & $6.426$ & $7.8\times10^{2}c$\\
$H$ & $1.29\times 10^{16}$ & $4.30\times 10^{5}$ & $1.00\times10^{3}$ & $6.309$ & $4.3\times10^{2}c$\\
$I$ & $2.67\times 10^{16}$ & $8.99\times 10^{5}$ & $1.00\times10^{4}$ & $4.101$ & $89c$\\
$J$ & $4.31\times 10^{16}$ & $1.49\times 10^{6}$ & $5.00\times10^{4}$ & $2.097$ & $29c$\\
\end{tabular}
\end{table}

\section*{THE GRB~980425 LUMINOSITY IN SELECTED ENERGY BANDS PREDICTED BY THE EMBH MODEL}

Recently, within the EMBH model, we have developed an attempt to theoretically derive the GRB spectra out of first principles as well as the GRB luminosity in fixed energy bands (Ruffini et al. 2003c). We have adopted three basic assumptions: {\bf a)} the resulting radiation as viewed in the comoving frame during the afterglow phase has a thermal spectrum and {\bf b)} the ISM swept up by the front of the shock wave, with a Lorentz gamma factor between $300$ and $2$, is responsible for this thermal emission. {\bf c)} We also adopt, like in our previous papers (Ruffini et al., 2001a,2001b,2002,2003b), that the expansion occurs with spherical symmetry. This three assumptions are different from the ones adopted in the GRB literature, in which the afterglow emission is believed to originate from synchrotron emission in the production of a shock or reverse shock generated when the assumed jet-like ejecta encounter the external medium (see e.g. Giblin et al., 2002 and references therein).

In the EMBH model the structure of the shock is determined by mass, momentum and energy conservation: the constancy of the specific enthalpy, which is a standard condition in shock rest frames (Zel'dovich \& Rayzer, 1966) and have been used in our derivation (Ruffini et al., 2003b). The only free parameter of our model is the size of the ``effective emitting area'' in the shock wave front: $A_{eff}$. Since the determination of this free parameter is performed here by empirically fitting the observational data, we avoid ambiguities due to the absence of relevant theoretical and laboratory results on relativistic shocks for Lorentz factor $\gamma\sim 300$.

The temperature $T$ of the black body in the comoving frame is then
\begin{equation}
T=\left(\frac{\Delta E_{\rm int}}{4\pi r^2 \Delta \tau \sigma {\cal R}}\right)^{1/4}\, ,
\label{tcom}
\end{equation}
where
\begin{equation}
{\cal R}=\frac{A_{eff}}{A_{abm}}
\label{rdef}
\end{equation}
is the ratio between the ``effective emitting area'' and the ABM pulse surface $A_{abm}$, $\sigma$ is the Stefan-Boltzmann constant and $\Delta E_{\rm int}$ is the proper internal energy developed in the collision between the ABM pulse and the ISM in a proper time interval $\Delta \tau$ (see Ruffini et al., 2003b,2003c). The ratio ${\cal R}$, which is a priori a function that varies as the system evolves, is evaluated at every given value of the laboratory time $t$.

All the subsequent steps are now uniquely determined by the equations of motion of the system. The basic tool in this calculation involves the definition of the EQuiTemporal Surfaces (EQTS) for the relativistic expanding ABM pulse as seen by an asymptotic observer. The key to determining such EQTS (see Fig.~1 in Ruffini et al., 2002) is the relation between the time $t$ in the laboratory frame at which a photon is emitted from the ABM pulse external surface and the arrival time $t_a^d$ at which it reaches the detector. We have instead adopted the equations (see Ruffini et al., 2002,2003b):
\begin{eqnarray}
t_a^d  =\left(1+z\right)\left(t-\frac{r\left(t\right)}{c}\cos\vartheta+\frac{{r_0 }}{c}\right)
=\left(1+z\right)\left(t - \frac{{\int_0^t {v\left( {t'} \right)dt'}  + r_0 }}{c}\cos \vartheta  + \frac{{r_0 }}{c}\right)
\nonumber \\ 
=\left(1+z\right)\left(t-\cos\vartheta\int_0^t {\sqrt{1-\frac{1}{\gamma^2\left(t'\right)}}dt'}+\frac{r_0}{c}\left(1-\cos\vartheta\right)\right)\, ,
\label{tad_fin}
\end{eqnarray}
where $z$ is the cosmological redshift of the source, $r_0\equiv r\left(t=0\right)$ and $\vartheta$ is the angle subtended by the emission point of the photon on the ABM pulse external surface, having defined $\vartheta=0$ along the line of sight. For such a relation, different approximations exist in the literature (see e.g. Fenimore et al., 1996; Sari, 1997,1998, Rees \& M\'esz\'aros, 1998; Fenimore et al., 1999; Granot et al., 1999 and see also the reviews by Piran, 1999; M\'esz\'aros, 2002). There are both quantitative and qualitative important differences in the use of Eq.(3) instead of the ones usually adopted in the literature (Ruffini et al., 2003b). It is a matter of fact that only the use of the EQTS defined by Eq.(3) allows to fit the observational data and that even a very minor departure from it leads to unacceptable results. It is also important to recall that this inadequacy in the current literature in the relation between the arrival time and the laboratory time has affected also all the estimates of the power-law slopes in the afterglow, preempting all current theoretical considerations in the literature of their predictive power. All the considerations about beaming in GRBs existing in the current literature have to be reformulated on the ground of the proper theoretical treatment and of Eq.(3) (see Ruffini et al., 2003b). Having so determined the EQTS we have been able to evaluate the source luminosity in a given energy band, in agreement with the above mentioned new assumptions (Ruffini et al., 2003c).

We can now proceed to the best fit of the GRB~980425 observed data. The best fit of the observed luminosity in selected energy bands has been obtained for the above mentioned values of the parameters $E_{dya}$, $B$ and for ${\cal R}$ monotonically decreasing from $4.81\times 10^{-10}$ to $2.65\times 10^{-12}$. The results are given in Fig.~2 where the luminosity is computed as a function of the arrival time for three selected energy bands.

\begin{figure}
\centering
\includegraphics[width=59mm]{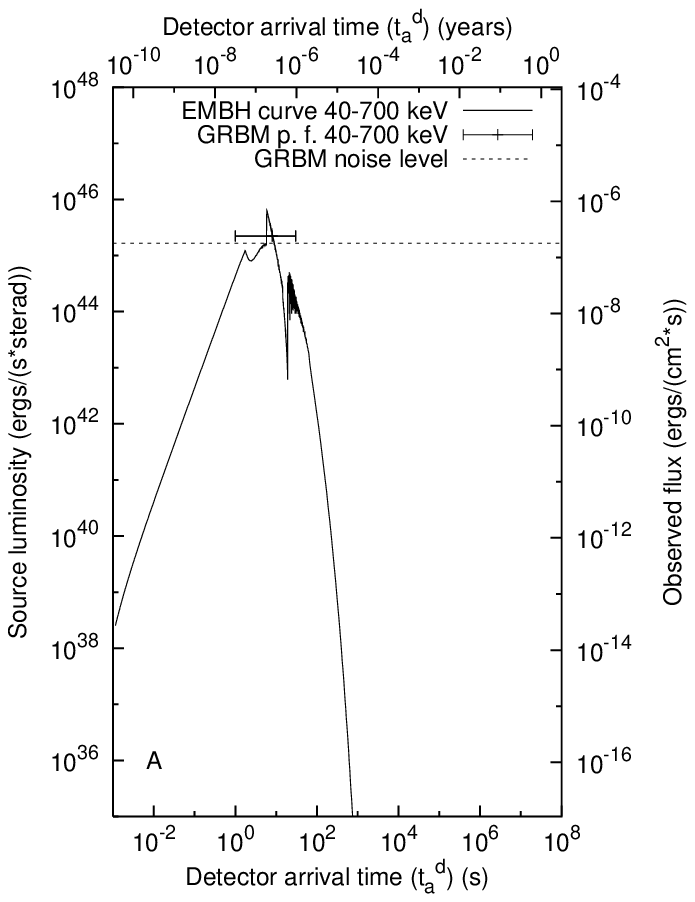}
\includegraphics[width=59mm]{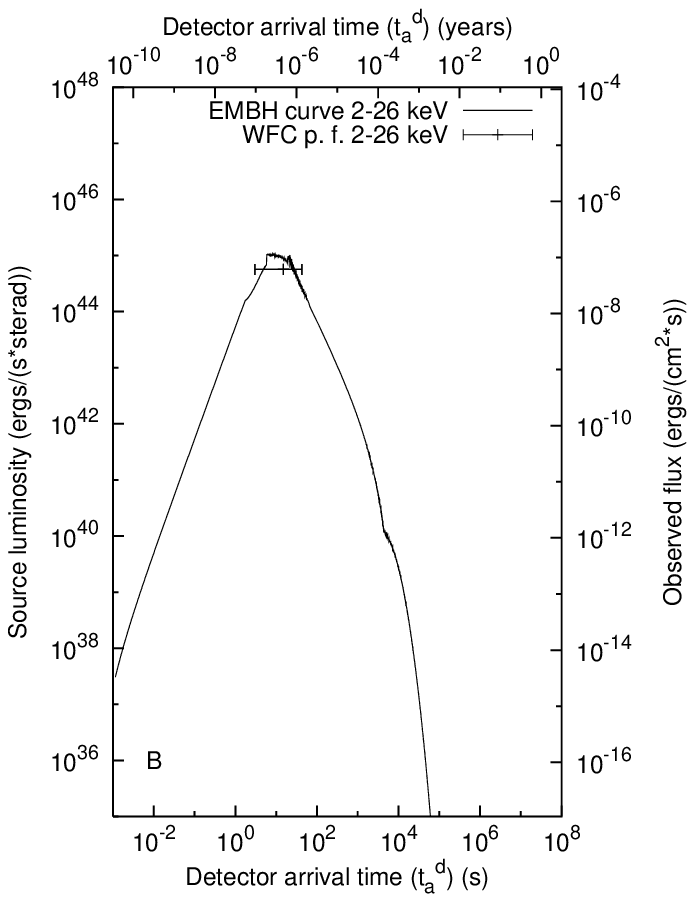}
\includegraphics[width=59mm]{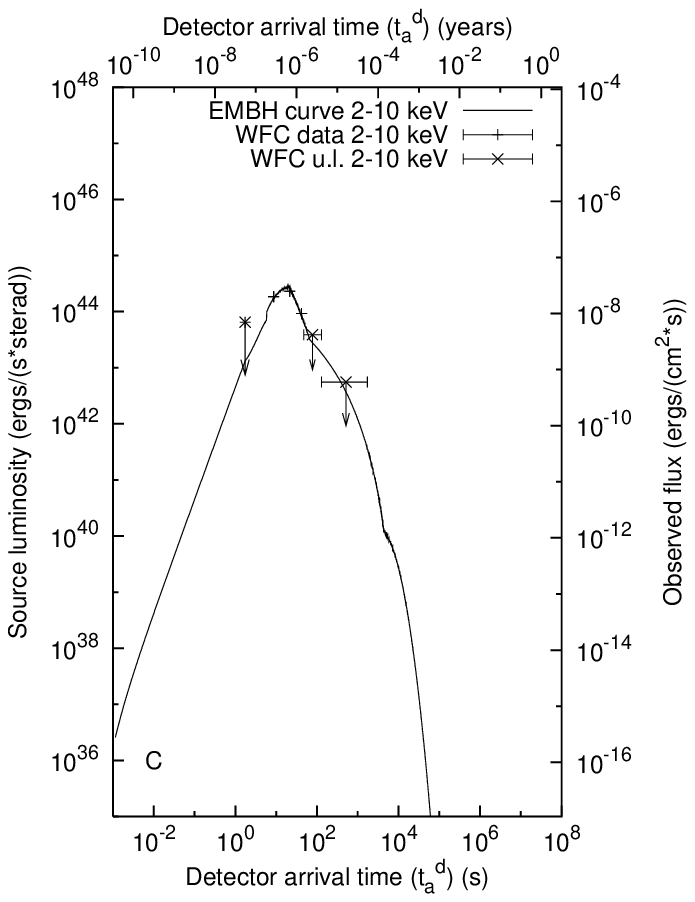}
\caption{{\bf a)} The light curve computed within the EMBH model in $\gamma$-ray (40-700 keV) is represented; the horizontal bar represents the peak flux in the 40--700 keV band measured by the GRBM (Frontera et al., 2000). The horizontal dotted line represents the noise level of the GRBM detector. {\bf b)} The light curve computed within the EMBH model in hard X-rays in the 2--26 keV band with the peak flux and time duration from WFC (Frontera et al., 2000). {\bf c)} The light curve computed within the EMBH model in hard X-rays in the 2--10 keV band with the experimental data in the same band from WFC (Pian et al., 2000)}
\end{figure}

We can then conclude:
\begin{enumerate}
\item The best fit is obtainable under almost perfect spherical symmetry. This has been proven as a result of an analysis of unprecedented redundancy: the luminosity curves are obtained from an integration over almost $10^7$ different paths, relating the observer to the EQTS. This procedure tests, to a very high level of accuracy, any departure from spherical symmetry as well as any departure from the computed equations of motion of the source (Ruffini et al., 2003b). This same circumstance was encountered in GRB~991216 (Ruffini et al., 2003b).
\item Each luminosity curve as a function of the arrival time presents complex behavior, which could be erroneously interpreted as evidence for breaks in the power-law indexes leading to erroneous inferences on the possible existence of jets (Ruffini et al., 2003c).
\item Although each luminosity curve presents some special features, the bolometric luminosity has a very clear and simple power-law behavior with the ``golden value'' index $n=-1.6$ (Ruffini et al., 2003b).
\end{enumerate}

\section*{THE NEW ASTROPHYSICAL SCENARIO AND THE NEWLY BORN NEUTRON STAR}

In Fig.~3 the luminosities in the three bands are represented together with the optical data of SN1998bw (black dots), the source S1 (black squares) and the source S2 (open circles). It is then clear that GRB~980425 is separated both from the supernova data and from the sources S1 and S2.

\begin{figure}
\centering
\includegraphics[width=\hsize]{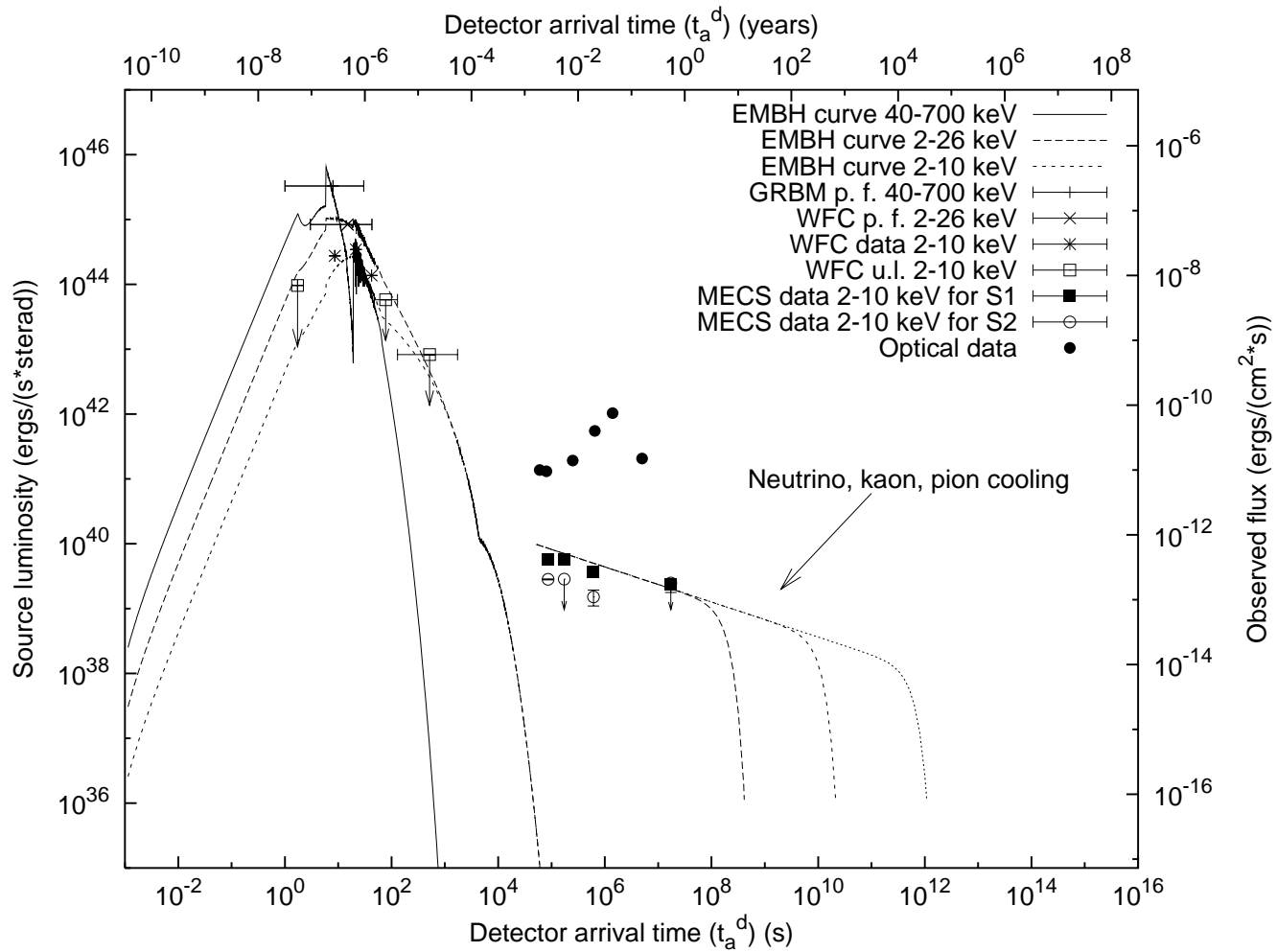}
\caption{The light curves in selected bands are reported as well as the MECS light curves in the 2-10 keV band of S1 and S2 (Pian et al. 2000) as well as the optical data (Iwamoto, 1999). Here are also reported theoretical models of neutron star cooling (Canuto, 1978).}
\end{figure}

While the occurrence of the supernova in relation to the GRB has already been discussed with the GRB-Supernova Time Sequence (GSTS) paradigm (Ruffini et al. 2001c), we like to address here a different fundamental issue: the possibility of observing the birth of a newly formed neutron star out of the supernova event, which in turn has been triggered by the GRB~980425.

In the early days of neutron star physics it was clearly shown by Gamow \& Schoenberg (1940,1941) that the URCA processes are at the very heart of the supernova explosions. The neutrino-antineutrino emission described in the URCA process is the essential cooling mechanism necessary for the occurrence of the process of gravitational collapse of the imploding core. Since then, it has become clear that the newly formed neutron star can be still significantly hot and in its early stages will be associated to three major radiating processes (Tsuruta, 1964,1979,2002; Canuto, 1978): {\bf a)} the thermal radiation from the surface, {\bf b)} the radiation due to neutrino, kaon, pion cooling, and {\bf c)} the possible influence in both these processes of the superfluid nature of the supra-nuclear density neutron gas. Qualitative representative curves for these cooling processes, which are still today very undetermined due to the lack of observational data, are shown in Fig.~3.

It is of paramount importance to follow the further time history of the two sources S1 and S2. If, as we propose, S2 is a background source, its flux should be practically constant in time and this source has nothing to do with the GRB~980425~/~SN1998bw system. If S1 is indeed the cooling radiation emitted by the newly born neutron star, it should be possible to notice a very drastic behavior in its luminosity as qualitatively expresses in Fig.~3.

\section*{CONCLUSIONS}

It is particularly attractive, in conclusion, to emphasize some of the analogies and the differences between the case of GRB~991216 and the one of GRB~980425:
\begin{enumerate}
\item In both these sources a GRB and, independently, a supernova event are present. In the case of GRB~991216 the inferences of the supernova can be obtained only on the ground of the emission of iron lines (Piro et al., 2000; Ruffini et al., 2001c). In the present case of GRB~980425 we have a very fortunate circumstance: the GRB source is much weaker and is much closer to us ($z=1.0$ for GRB~991216 and $z=0.00835$ for GRB~980425). This situation is particularly important for obtaining detailed data on the supernova and on the possible occurrence of a newly born neutron star. This occurrence could not be observed in GRB~991216 due to the very large distance and to the overwhelming X-ray luminosity of the afterglow (see Ruffini et al., 2003b).
\item The energetics of GRB~991216 is $E_{dya}=4.83\times 10^{53}\, {\rm ergs}$, while the one of GRB~980425 is $E_{dya}=1.1\times 10^{48}\, {\rm ergs}$. It is very impressive that the EMBH model applies over a range of more than 6 orders of magnitude, giving important inferences on the source as well as on the structure of the ISM surrounding the source. It is significant that in both sources the condition of spherical symmetry appears to be strongly implemented and this fact is a very clear discriminant among all possible sources of energy for GRBs.
\item The main difference between GRB~991216 and GRB~980425 is then traced back, within the EMBH model, to the different parameters occurring in the dyadosphere and to the nature of the ``effective'' Reissner-Nordstr\"om geometry which we have used as a reliable estimate for the dynamical processes of gravitational collapse leading to the formation of the EMBH. For GRB~991216 we have a ratio between the electromagnetic energy and the total mass of the imploding core, described by the parameter $Q/M$, given by $Q/M=0.23$ while for GRB~980425 we have $Q/M=6.5\times 10^{-4}$. In both cases a reasonable mass of the black hole is $M_{BH}\simeq 10M_{\odot}$ (Ruffini et al., 2003b).
\end{enumerate}

A dedicated observational campaign, both with XMM and Chandra, to follow the cooling of the newly formed neutron star is needed in order to gain for the first time information on this extremely important astrophysical process.

\end{document}